\documentclass[reprint,showpacs,amsmath,amssymb,superscriptaddress]{revtex4-2}
\usepackage{graphicx}
\usepackage{dcolumn}
\usepackage{ulem}
\usepackage{array}
\usepackage{color}
\usepackage{multirow}

\usepackage[usenames,dvipsnames]{xcolor} 
\usepackage{changes}

\begin{document}

\title{Numerical investigation of a segmented-blade ion trap with biasing rods}

\author{Jungsoo Hong}
\author{Myunghun Kim}
\author{Hyegoo Lee}
\author{Moonjoo Lee}
\email{moonjoo.lee@postech.ac.kr}
\affiliation{Department of Electrical Engineering, Pohang University of Science and Technology (POSTECH), 37673 Pohang, Korea}

\newcommand{\new}[1]{\textcolor{red}{#1}}

\date{\today}

\begin{abstract}
We report a numerical study of a linear ion trap that has segmented blades and biasing rods. 
Our system consists of radio frequency (rf) blades, dc blades with ten separate electrodes, and two biasing rods for compensating the ions' micromotion.
After calculating the optical access for the ions, we find rf and dc voltages that result in a stable trapping configuration of $^{171}$Yb$^{+}$ ions. 
We also explore the micromotion compensation with the biasing rods, and calculate the influence of blade misalignment to the trap potential. 
Our work offers quantitative understanding of the trap architecture, assisting reliable operation of an ion-trap quantum computer.
\end{abstract}

\maketitle

\section{Introduction}

Trapped atomic ions are one of the best-controlled systems for quantum information processing~\cite{Leibfried03, Blatt08, Monroe13}. 
The ions are typically stored from days to weeks in an ultrahigh vacuum environment.
The record infidelity is as low as 10$^{-6}$ for single-qubit gates~\cite{Harty14} and 10$^{-3}$ for two-qubit gates~\cite{Ballance2016, Gaebler2016}, while the state preparation and measurement errors are below 10$^{-4}$~\cite{Harty14, Crain2019}.
The ion qubit's coherence time is longer than one hour in optimized conditions~\cite{Wang2017, Wang2021}.
Quantum entanglement of 20 ions has been achieved through addressing the individual ions~\cite{Friis2018}, and even entanglement of 24 ions was demonstrated in a very compact setting~\cite{Pogorelov2021}.
Moreover, the ``full connectivity'', mediated by collective motion of the ions, makes it possible to execute quantum algorithms in efficient ways~\cite{Wright2019}.

Among a number of ion-trap architectures~\cite{Siverns2017a, Romaszko2020}, a macroscopic trap with thin segmented blades~\cite{Hucul2015a} would be an excellent setting for quantum experiments.  
This trap contains two diagonal rf blades in order to generate a psuedopotential for confining the ions along the transverse directions.
The axial potential is formed by two dc blades aligned along the anti-diagonal direction, via applying dc voltages to each of the segmented electrodes. 
Owing to the comparatively smaller thickness of the blades than those of e.g., Refs.~\cite{Huber2008, Schnitzler2009}, the interblade distance could become short, which enables the generation of deep trapping potentials. 
Seminal experiments with this trap were reported, including the realization of a modular quantum network~\cite{Hucul2015}, small programmable quantum computer~\cite{Debnath2016}, and global or parallel entangling gates~\cite{Lu2019, Figgatt2019}.

We point out three major advantages of the segmented-blade trap.
First, the trap offers a great optical access for the qubit control and measurement~\cite{Hucul2015, Siverns2017}.
It is possible to address the ions from radial directions with a numerical aperture (NA) as large as 0.6~\cite{Hucul2015}.
Second, the trap provides the capability of engineering various potential shapes along the trap axis.
Adjusting dc voltages to each segment, the trap can create harmonic, quartic, and multi-well potentials~\cite{Lin2009}.
Lastly, this trap exhibited great performance at cryogenic temperatures: Recent experiment demonstrated the trapping of more than 100~ions in a cryogenic environment~\cite{Pagano2018}, representing the scalability of qubit numbers in this platform.


Here, we present a comprehensive numerical study on this segmented-blade ion trap.
In addition to the structure of Ref.~\cite{Hucul2015a}, we include two biasing rods~\cite{berkeland98, McLoughlin2011, Schindler13, Xie2021, Kim2022} solely for compensating the micromotion. 
We investigate the optical access, and calculate the potential energy for trapping $^{171}$Yb$^{+}$ ions. 
Our calculation shows that the micromotion can be compensated by applying dc voltages to the biasing rods and dc electrodes.
We also describe the influence of blade misalignment to the trap potential.

\begin{figure*} [!t]
	\includegraphics[width=6.3in]{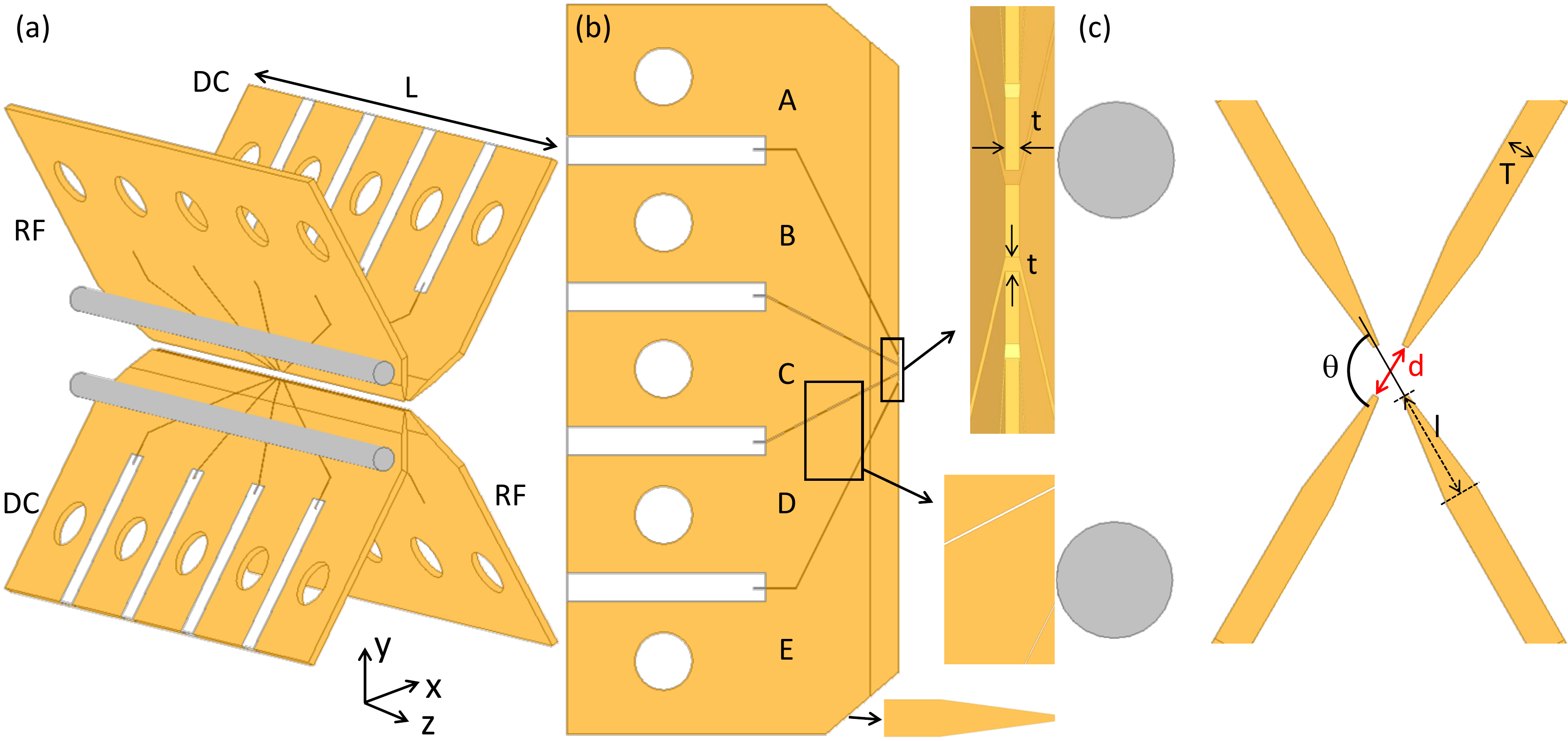} 
	\caption{
	(a) Overview of segmented blade trap.	
	Length of the blade $L$ is 25.2~mm, and diameter of the biasing rod (gray) is 1~mm.
	(b) Top view of dc blade. 
	Zone in gold denotes metal-coated facet; white rectangles indicate insulating ceramic surface. 	
	Segments from A to E are disconnected electrically. 
	Holes are used to insert screws for fixing the blade to mount (not shown).
	Width of tapered end and groove, marked as $t$, is 50~$\mu$m.
	(c) Central area of the trap in the $xy$ plane.
	Thickness of the blade, referred to as $T$, is 300~$\mu$m; tapered length $l$ is 1~mm; angle between the blades $\theta$ is 120$^{\circ}$.
	Blade-blade distance is given by $d$.
	}
	\label{fig:design}
\end{figure*}

\begin{figure} [!b]
	\includegraphics[width=3.3in]{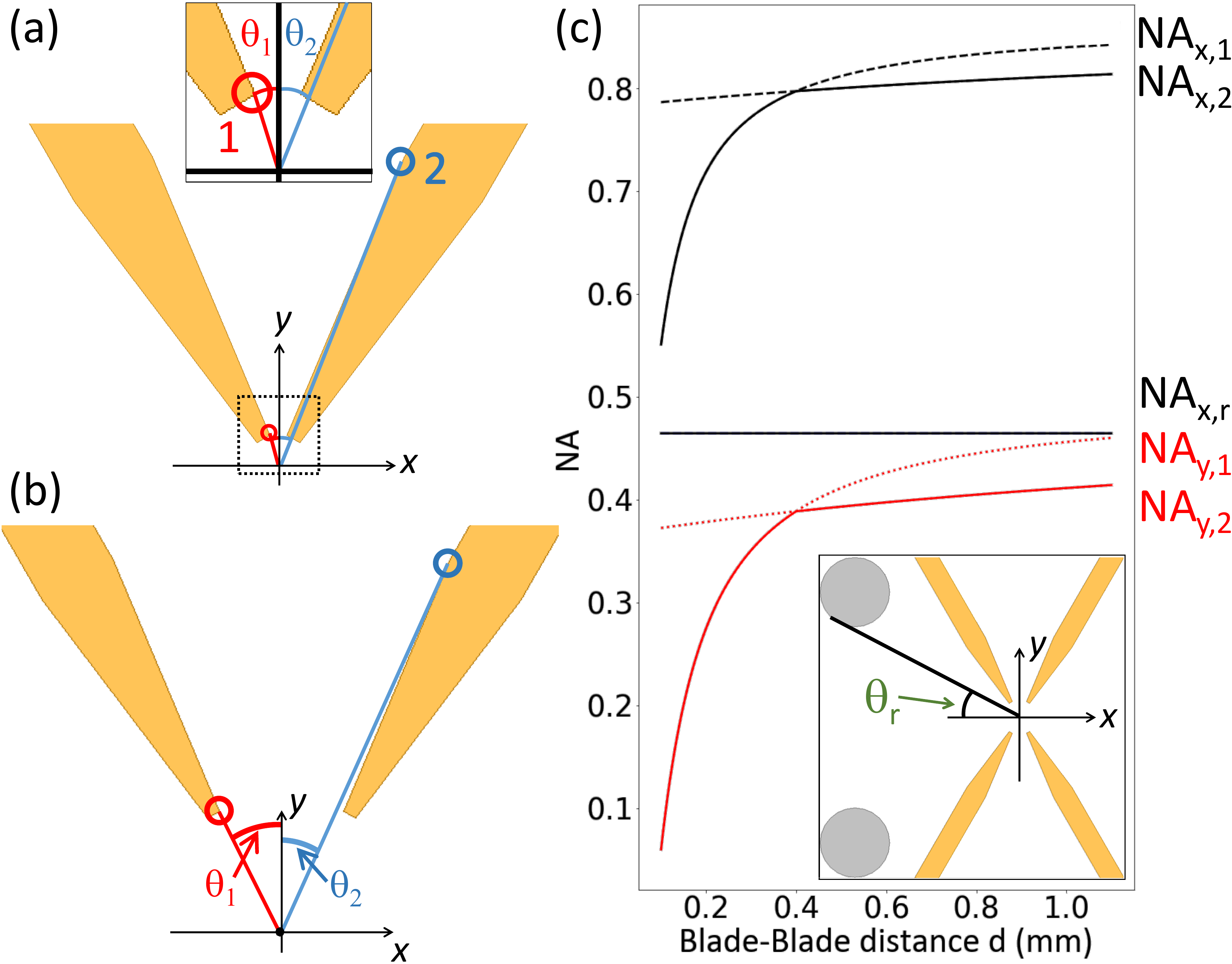}
	\caption{
	(a) Configuration when blade-blade distance $d < d_{0}$.
	Red circle corresponds to point 1 and blue to point 2.
	Available NA is determined by $\theta_{1}(<\theta_{2})$.
	(inset) Zoomed-in view of central area.
	(b) Actual NA is limited by  $\theta_{2}(<\theta_{1})$ at $d > d_{0}$.
	The angle $\theta_{1}$ coincides with $\theta_{2}$ at $d=d_{0}$.
	(c) NAs are plotted as a function of $d$, along $x$ direction in black and $y$ direction in red. 	
	NA$_{x/y, 1/2/\rm{r}}$ means the NA limited along $x/y$ axis by the point $1, 2$, and rod. 
	Actual NAs are indicated by solid lines.
	NA$_{x, 1/2}$ works for $x>0$, and NA$_{x, \rm{r}}$ for $x<0$.
	Inset shows $\theta_{\rm{r}}$.
	}
	\label{fig:NA}
\end{figure}

\section{Result}

\subsection{Trap design}

As presented in Fig.~\ref{fig:design}(a), we consider four segmented blades (gold) in a linear trap configuration, with two biasing rods (gray). 
Two diagonal blades are driven with a rf voltage, and the anti-diagonal ones with ten different dc voltages.
While all five segments in the rf blade are electrically connected, the dc blades' segments are disconnected by the grooves and nonmetal zones (white area in Figs.~\ref{fig:design}(a) and (b)).
We can apply different dc voltages to the zones from A to E.  
The trap potential along the $xy$ directions is mostly derived from the rf voltage with minor contribution from the dc segments; the potential along the $z$ direction originates from the voltages to all the dc electrodes.
Fig.~\ref{fig:design}(c) shows a diagonal blade-blade distance of $d$ at an angle $\theta=120^{\circ}$ between the rf and dc blades. 

\begin{figure*} [!t]
	\includegraphics[width=6.3in]{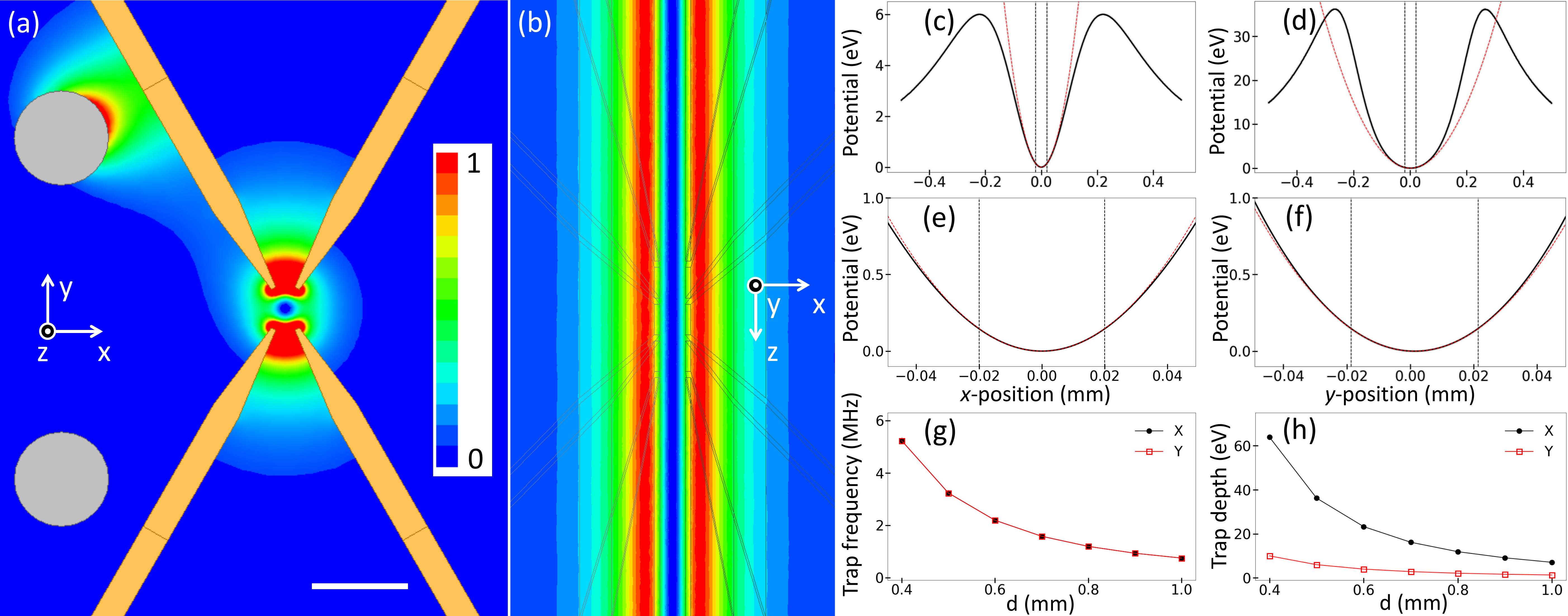}
	\caption{
	Contour plot of calculated pseudopotential on $xy$ plane at $z=0$ (center of trap) in (a), and $xz$ plane at $y=0$ in (b), with $V_{\rm{rf}}=600$~V.
	Scale bar indicates 1~mm.
	Slicecut of the trap potential is shown with black lines along $x$ axis in (c), and $y$ axis in (d).
	Central regions are focused in (e) and (f).
	Black solid line is result of the numerical simulation; red dashed line is fit to a quadratic function; vertical dashed lines indicate fitting range.
	(g), (h) Trap frequency and depth along $x$ and $y$ directions as a function of blade-blade distance $d$.
	}
	\label{fig:pseudopotential}
\end{figure*}

\subsection{Optical access}

We discuss the optical access for measuring the ion fluorescence and addressing individual ions.
Figs.~\ref{fig:NA}(a) and (b) show that the two tapering corners 1 and 2 determine available NAs. 
When the blade-blade distance $d$ is smaller than $d_{0} = 2  l\times t/(T-t)=400$~$\mu$m, we find that $\theta_{1}$ is smaller than $\theta_{2}$, resulting in a situation where the NA is limited by $\theta_{1}$.
The size of the two angles is reversed at $d > d_{0}$ so that the NA is determined by $\theta_{2}$ in this regime; the two angles become identical for the case of $d=d_{0}$.
Considering the trap geometry, we derive analytical formulas for the NAs along the $x$ and $y$ directions. When $d<400$~$\mu$m, we obtain
\begin{align*}
	\textrm{NA}_{x, 1} &= \sin{ \left(   \frac{\pi}{3} - \arctan{ \left( \frac{t}{d}   \right)}         \right)   } \\ 	
	\textrm{NA}_{y, 1} &= \sin{ \left(   \frac{\pi}{6} - \arctan{ \left( \frac{t}{d}   \right)}         \right)   },  \\ 	
\end{align*}
\noindent
and the expressions below describe the NAs at $d>400$~$\mu$m,
\begin{align*}
	\textrm{NA}_{x, 2} &= \sin{ \left(   \frac{\pi}{3} - \arctan{ \left( \frac{T}{2l+d}   \right)}        \right)   } \\  
	\textrm{NA}_{y, 2} &= \sin{ \left(   \frac{\pi}{6} - \arctan{ \left( \frac{T}{2l+d}   \right)}        \right)   }, \\ 	
\end{align*}

\noindent
with the thickness of the tapered end $t$, tapered length $l$, and untapered thickness of the blade $T$. 
The NA determined by the point $i$ along the $x/y$ axis is denoted by NA$_{x/y, i}$~$(i=1, 2)$ plotted in Fig.~\ref{fig:NA}(c).
The actual NA corresponds to the smaller value between the two NAs at a given distance. 
For example, we attain NA$_{x}=0.77$ and  NA$_{y}=0.35$ at $d=300$~$\mu$m that is limited by the point 1, while at $d=500$~$\mu$m, the NA$_{x}$ increases to 0.80 and NA$_{y}$ to 0.39 determined by the point 2.
It is notable that, for the direction $x<0$, the optical access is limited by the biasing rods $\theta_{\rm{r}}$ with NA$=0.46$.

\subsection{Simulation method}

The total trap potential $\phi(\bold{r})_{\rm{tot}}$ is given by the sum of the pseudopotential generated by a rf voltage applied to the two rf blades, static potential $\phi(\bold{r})_{\rm{dc}}$ derived from dc voltages on each segment of the dc blades, and another static potential $\phi(\bold{r})_{\rm{rod}}$ from biasing rods:

\begin{align}
	\phi(\bold{r})_{\rm{tot}} & = \phi(\bold{r})_{\rm{rf}} + \phi(\bold{r})_{\rm{dc}} + \phi(\bold{r})_{\rm{rod}}\nonumber \\ 
							   & = \frac{q^2}{4m\Omega^2} |\vec{E}(\bold{r})|^2  + q  \sum_{i} V(\bold{r})_{i} + q V(\textbf{r} )_{\rm{rod}}, \nonumber
\end{align}

\noindent
where $q$ and $m$ represent the charge and mass of a single ion, respectively and $\vec{E}(\bold{r})$ refers to the electric field stemmed from the rf voltage, $V(\bold{r})_{i}$ is the voltage derived from the dc electrode $i$, and $V(\textbf{r} )_{\rm{rod}}$ is the dc voltage from biasing rods.
We choose the driving frequency $\Omega = 2\pi\times 22.5$~MHz in this paper. 
All calculations are done for $^{171}$Yb$^{+}$ ions using the software ANSYS Electronics Desktop 2020 R1.

\subsection{Pseudopotential}
\label{sec:Pseudopotential}

Our simulation results of the rf pseudopotential are presented in Fig.~\ref{fig:pseudopotential}. 
We begin with the case of $d=500$~$\mu$m.
Both rf blades are driven with an amplitude of $V_{\rm{rf}}=600$~V, and all dc segments and biasing rods are ground. 
In this configuration, as shown in Figs.~\ref{fig:pseudopotential}(a) and (b), a pseudopotential ``tube'' is formed along the $z$ axis.
We then obtain the trap frequency $\omega_{r}/2\pi$ via fitting the potential to a quadratic function $\phi(r)=m\omega_{r}^2 r^2/2$, for $r=x$ and $y$.
The fitting range is limited to the central region (from $-20$ to $+20~\mu$m with respect to the potential minimum, marked in Figs.~\ref{fig:pseudopotential}(c)-(f)) where the contribution of higher order terms is negligible.  
The yielded trap frequencies are 3.2~MHz along both the $x$ and $y$ axes. 
Concerning the trap depth $U_{0}$, we define the depth as the maximum value of the potential energy in a given direction: We obtain 6.0~eV along the $x$ axis and 36~eV to the $y$ direction.

\begin{figure} [!b]
	\includegraphics[width=3.3in]{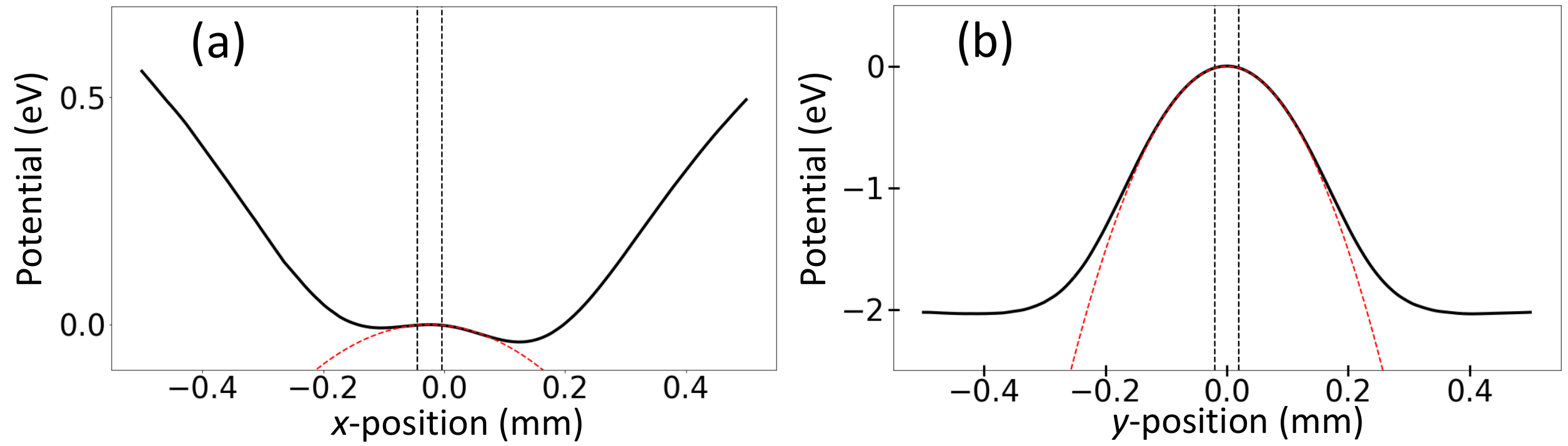}
	\caption{
	Static potential $\phi(\textbf{r})_{\rm{dc}}$ derived from dc voltages of $V_{\rm{A}} = V_{\rm{E}}=15$~V, $V_{\rm{B}} = V_{\rm{D}} = 3$~V,  $V_{\rm{C}}=-10$~V, along $x$ axis in (a) and $y$ axis in (b) at $z=0$. 
	Black solid line is the simulation result; red dashed line is fit to a quadratic function; the fitting range is defined by vertical dashed lines. 
	}
	\label{fig:dc_potential}
\end{figure}

\begin{figure*} [!t]
	\includegraphics[width=6.3in]{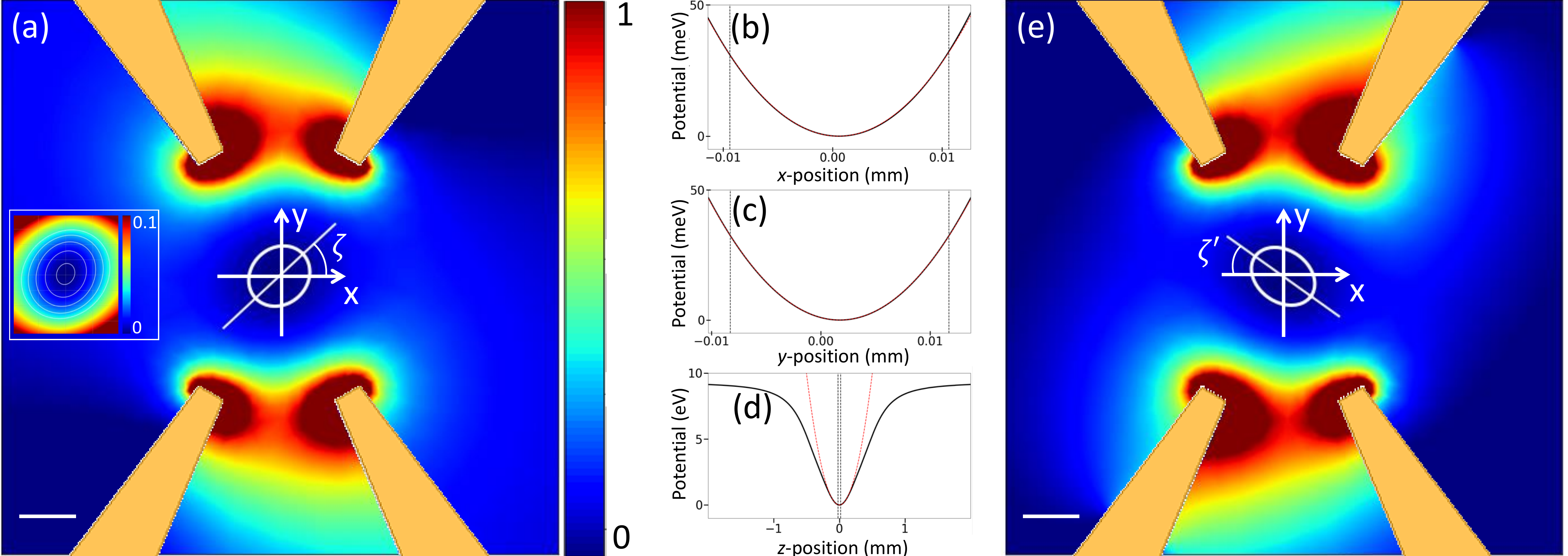}
	\caption{
	(a) Contour plot of total potential $\phi(x, y, 0)_{\rm{tot}}$ derived from rf and dc voltages given in main text. 
	Principal axis is rotated by $\zeta$ with respect to $x$ axis.
	Scale bar indicates 100~$\mu$m.
	Equipotential line along 1~eV is shown, and several lines in inset.
	Color scale is from 0 to 40~eV in (a) and 0 to 0.1~eV in inset.
	Total potential along $x$ axis in (b), $y$ axis in (c), and $z$ axis in (d). 
	Black solid line is simulation result; red dashed line is fit to quadratic function; fitting range is defined by vertical dashed lines. 
	(e) Contour plot of total trap potential $\phi(x, y, \pm1~\textrm{mm})_{\rm{tot}}$. 
	Principal axis is rotated by $180^{\circ}-\zeta'$ with respect to $x$ axis.
	}
	\label{fig:total_potential}
\end{figure*}

We remark two features of this simulation result.
First, the calculation shows that the ion confinement is stronger along the $y$~axis.
It is because the distance between the rf and dc blades is smaller than that of the $x$ direction, leading to a higher electric field magnitude along the $y$ axis. 
Second, as indicated in Figs.~\ref{fig:pseudopotential}(e) and (f), it is notable that the minimum point of the pseudopotential is shifted from the origin by (0.0, 1.4)~$\mu$m.
It is due to the fact that the potential energy, created between the biasing rod and rf blade at $x<0$ and $y>0$, shifts the rf-null point to $y>0$ slightly (see Appendix).
Near the origin, this potential  is approximated as $\phi(0, y)_{\rm{rf}}\simeq -c_{y} \cdot y$ ($c_{y}$ is positive). 
The total pseudopotential is then given by $\phi(0, y)_{\rm{rf}} =\frac{1}{2} m\omega_{y}^2( y - \frac{c_{y}}{m\omega_{y}^2})^2+const.$ along the $y$ axis, explaining the shifting direction of the potential minimum.

We continue such calculations for several $d$ (Figs.~\ref{fig:pseudopotential}(g) and (h)) at $V_{\rm{rf}}=600$~V.
Both the trap frequency and depth decrease as $d$ increases.
The trap depth is higher than several eV for all the distances, and the trap frequency decreases from 5.2~MHz to 0.7~MHz along both the $x$ and $y$ axes, as $d$ changes from 400~$\mu$m to 1~mm.

\subsection{Confinement in three dimensions}
\label{sec:confinement_in_three_dimensions}
  
The axial confinement of the ions is made by the static potential $\phi(\bold{r})_{\rm{dc}}$.
Here, we concentrate on the generation of a single-well, harmonic potential along the $z$ direction.
A positive dc voltage set of $V_{\rm{A}}=V_{\rm{E}}$ and $V_{\rm{B}}=V_{\rm{D}}$ gives rise to a stable symmetric potential along the $z$ axis.
However, due to Earnshaw's theorem~\cite{Steane1997}, the dc voltage set brings about a deconfinement effect along the radial directions (Fig.~\ref{fig:dc_potential}) --- the depth of pseudopotential is typically much larger than that of the deconfinement potential, which ensures the trapping of ions.
Overall, the sum of the static potential $\phi(\bold{r})_{\rm{dc}}$ and pseudopotential $\phi(\bold{r})_{\rm{rf}}$ provides a stable total trapping potential $\phi(\bold{r})_{\rm{tot}}$, which can be fitted with a quadratic function along all three directions.

In Fig.~\ref{fig:total_potential}, the total trap potential $\phi(\mathbf{r})_{\rm{tot}}$ is shown in our chosen voltage set of $V_{\rm{rf}}=600$~V, $V_{\rm{A, E}}=15$~V, $V_{\rm{B, D}}=3$~V, and $V_{\rm{C}}=-10$~V at $d=500$~$\mu$m.
The biasing rods are ground.
The potential is well fitted with a quadratic function along the three axes, and the fittings give the trap frequencies of $(\omega_x, \omega_y, \omega_z)/2\pi = (3.2, 3.1, 1.1)$~MHz for $^{171}$Yb$^{+}$ ion, with the depths of 6.1, 34, 9.1~eV at each direction, constituting a stable trapping configuration.
The potential minimum is located at $(0.2, 1.6, 0.0)$~$\mu$m.
Despite the deconfinement effect of the dc voltages in Fig.~\ref{fig:dc_potential}, the pseudopotential dominates the trapping potential along the $x$ and $y$ directions.
Note that the impression of a negative voltage to $V_{\rm{C}}$ offers both deeper axial potential and higher trap frequency. 
When $V_{\rm{C}} = 0$~V, the axial trap frequency $\omega_{z}/2\pi$ decreases to $0.63$~MHz and the trap depth along the $z$ axis to 6.5~eV.

It is noteworthy to mention that the principal axis is determined by the applied dc voltages.
In Fig.~\ref{fig:total_potential}(a), we plot a slicecut of $\phi(\textbf{r})_{\rm{tot}}$ at $z=0$.
In this region, the effect of dc voltage $V_{\rm{C}}=-10$~V governs the symmetry of the potential: The negative dc voltages reduce confinement along the both dc electrodes, which causes that the long principal axis aligns along the dc blades.
The angle between this axis and $x$ axis is $\zeta=35^{\circ}$ obtained by fitting the equipotential line with an ellipse. 
This phenomenon is reversed, as presented in Fig.~\ref{fig:total_potential}(e), at $z=\pm1~$mm where the effect of $V_{\rm{A/E}}=15$~V is maximized. 
The positive dc voltage increases the confinement along the dc blades, and thus the long principal axis of the equipotential line aligns along the rf blade. 
The principal axis rotates from $\zeta$ to $180^{\circ}-\zeta'=139^{\circ}$ continuously as $z$ coordinate changes from $z=0$ to $z=\pm1$~mm.

\begin{figure*} [!t]
	\includegraphics[width=6.3in]{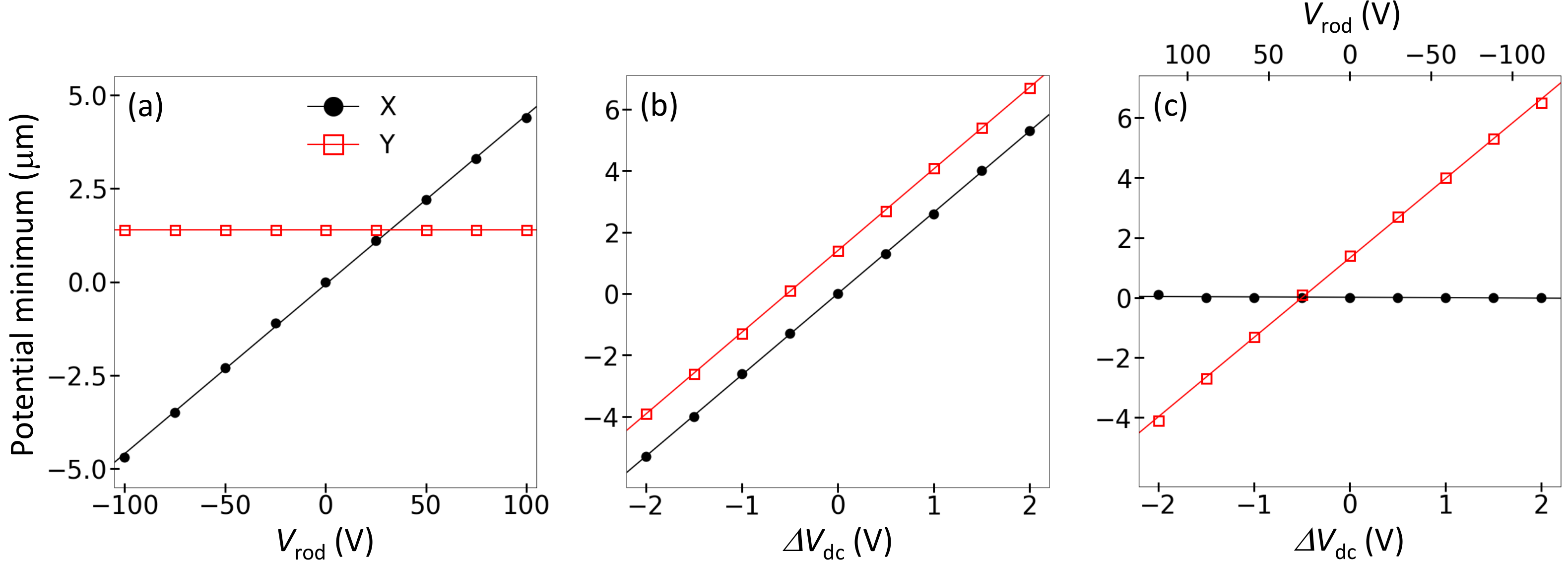}
	\caption{
	(a) Position of potential minimum as a function of $V_{\rm{rod}}$ with $\Delta V_{\rm{dc}}=0$.
	Solid lines are fit to linear functions.
	(b) Potential minimum point as $\Delta V_{\rm{dc}}$ changes at fixed $V_{\rm{rod}}=0$.
	(c) Position of potential minimum as $\Delta V_{\rm{dc}}$ changes like (b), however the position shift along the $x$ direction is canceled by varying $V_{\rm{rod}}$.
	Position error is dominated by spatial resolution of simulation, 100~nm (error bars not shown, similar with symbol size). 
	}
	\label{fig:micromotion}
\end{figure*}

\subsection{Micromotion compensation}
\label{sec:micromotion}

Next, we describe the simulation of micromotion compensation in our system. 
As discussed in Refs.~\cite{Debnath2016thesis, An2017thesis}, the micromotion could be compensated by adding independent and different dc offset voltages to each rf blade.
We approach this from a different perspective: Via including the biasing rods (gray in Fig.~\ref{fig:design}(a)), we have two independent control parameters, those rods and dc electrodes, so as to move the position of potential minimum arbitrarily in the transverse plane.

We first quantify the impact of biasing rods to the location of the potential minimum.
Applying $V_{\rm{rf}}=600$~V to rf blades with dc blades ground, we change the voltage on both the biasing rods $V_{\rm{rod}}$ from $-100$ to $100$~V, and search for the potential minimum of $\phi(\textbf{r})_{\rm{tot}}$. 
The positions of potential minimum are plotted in Fig.~\ref{fig:micromotion}(a): The location is only shifted along the $x$ axis over $\sim$10~$\mu$m. 
The linear fitting in Fig.~\ref{fig:micromotion}(a) gives a displacement per $V_{\rm{rod}}$ of $45.3(4)$~nm$/V_{\rm{rod}}$ along the $x$ direction. 
In terms of electric field, given $V_{\rm{rod}}=1$~V to both rods, an electric field of (33.3(1), 0.0)~V/m is generated at the origin, and at the rf-null position of $(0.0, 1.4)$~$\mu$m, slight electric field along the $y$ direction is obtained as $(33.3(1), 0.3(1))$~V/m.
The shift of the potential minimum induced by the small $y$ component is negligible along the $y$ axis.

\begin{figure*} [!t]
	\includegraphics[width=6.3in]{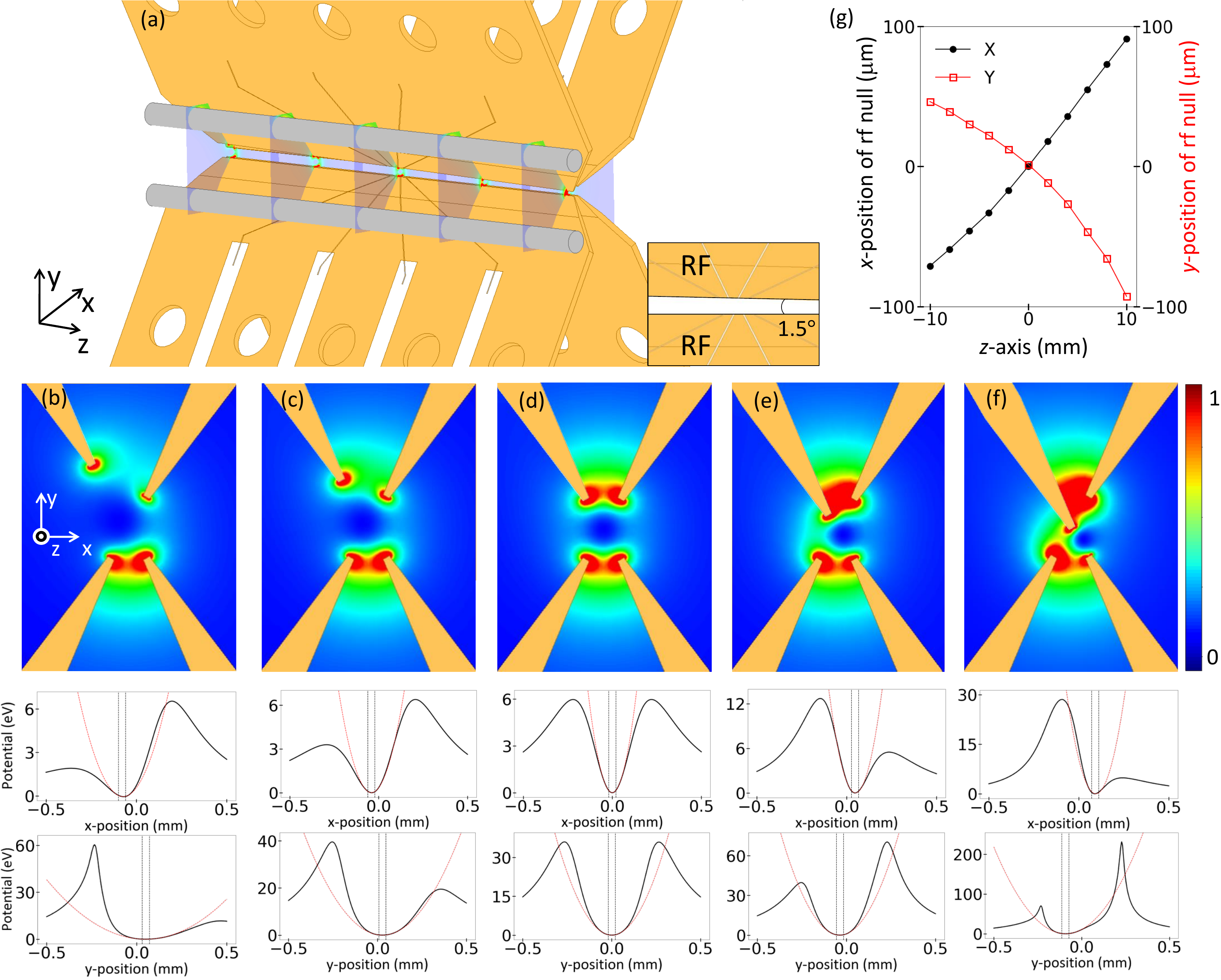}
	\caption{
	(a) Rf blade at $x<0, y>0$ is tilted by 1.5$^{\circ}$ with respect to the $z$ axis. 
	Semi-transparent planes correspond to the cross sections shown in (b)-(f).
	Contour plot of pseudopotential at $z=-10$~mm in (b), $z=-5$~mm in (c), $z=0$~mm in (d), $z=5$~mm in (e), and $z=10$~mm in (f). 
	Black solid line is the slicecut along the $x$ and $y$ axis, fitting range is defined by black dashed line, red dashed line is quadratic fitting.
	(g) Coordinates of rf null are plotted along the $z$ axis. 
 	}	
	\label{fig:tilt_xy_plane}
\end{figure*}

\begin{figure*} [!t]
	\includegraphics[width=6.0in]{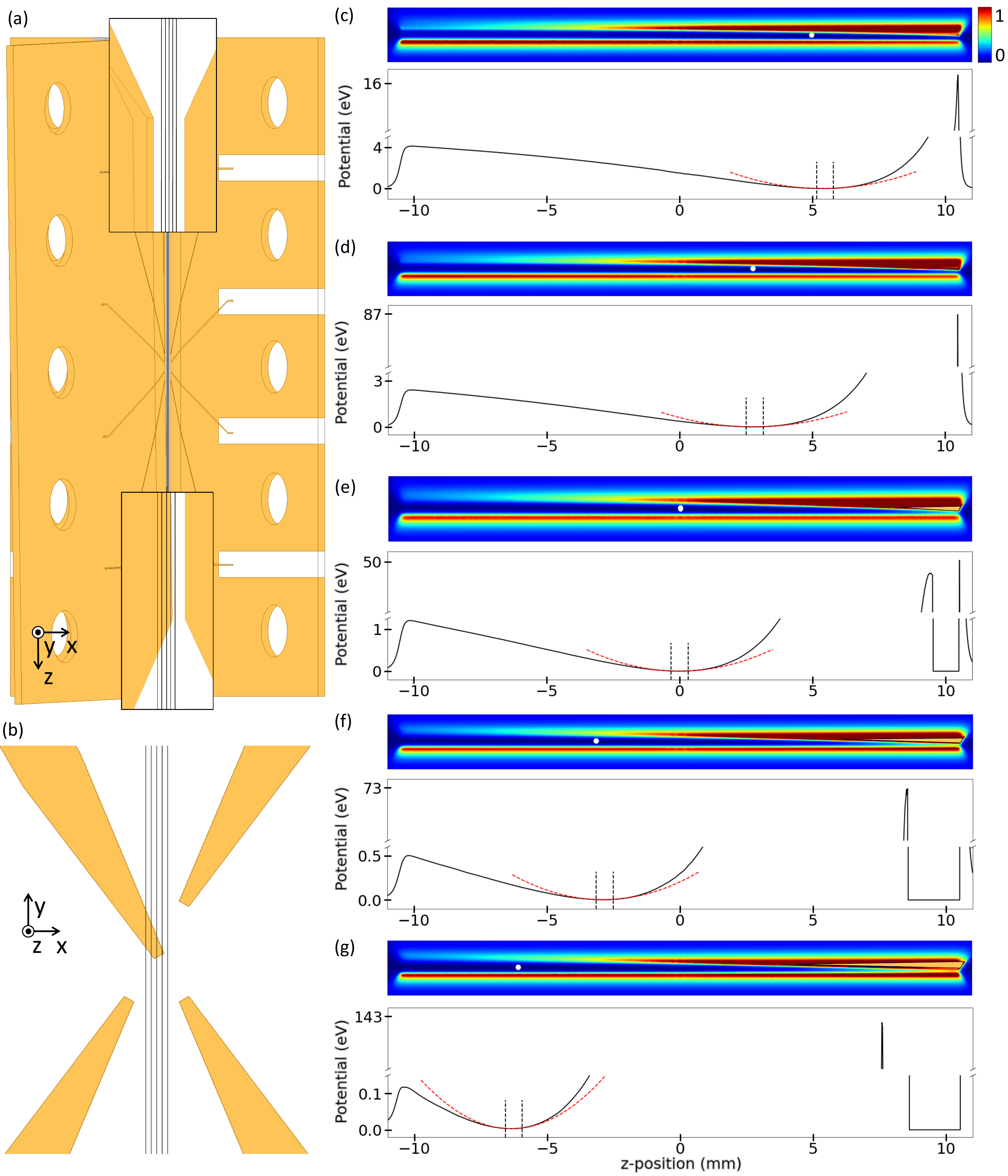}
	\caption{
	(a), (b) Cross sections of the pseudopotential in the $yz$ plane (black lines).
	Contour plot of the pseudopotential at $x=-50$~$\mu$m in (c), $x=-25$~$\mu$m in (d), $x=0$~$\mu$m in (e), $x=25$~$\mu$m in (f), and $x=50$~$\mu$m in (g).
	White dot indicates the minimum of pseudopotential.
	Black solid line is the slicecut along the $z$ axis at $y = 0$ of the given plane, black dashed line defines the fitting range, and red dashed line is quadratic fitting. 
	}	
	\label{fig:tilt_yz_plane}
\end{figure*}

\begin{figure} [h]
	\includegraphics[width=3.3in]{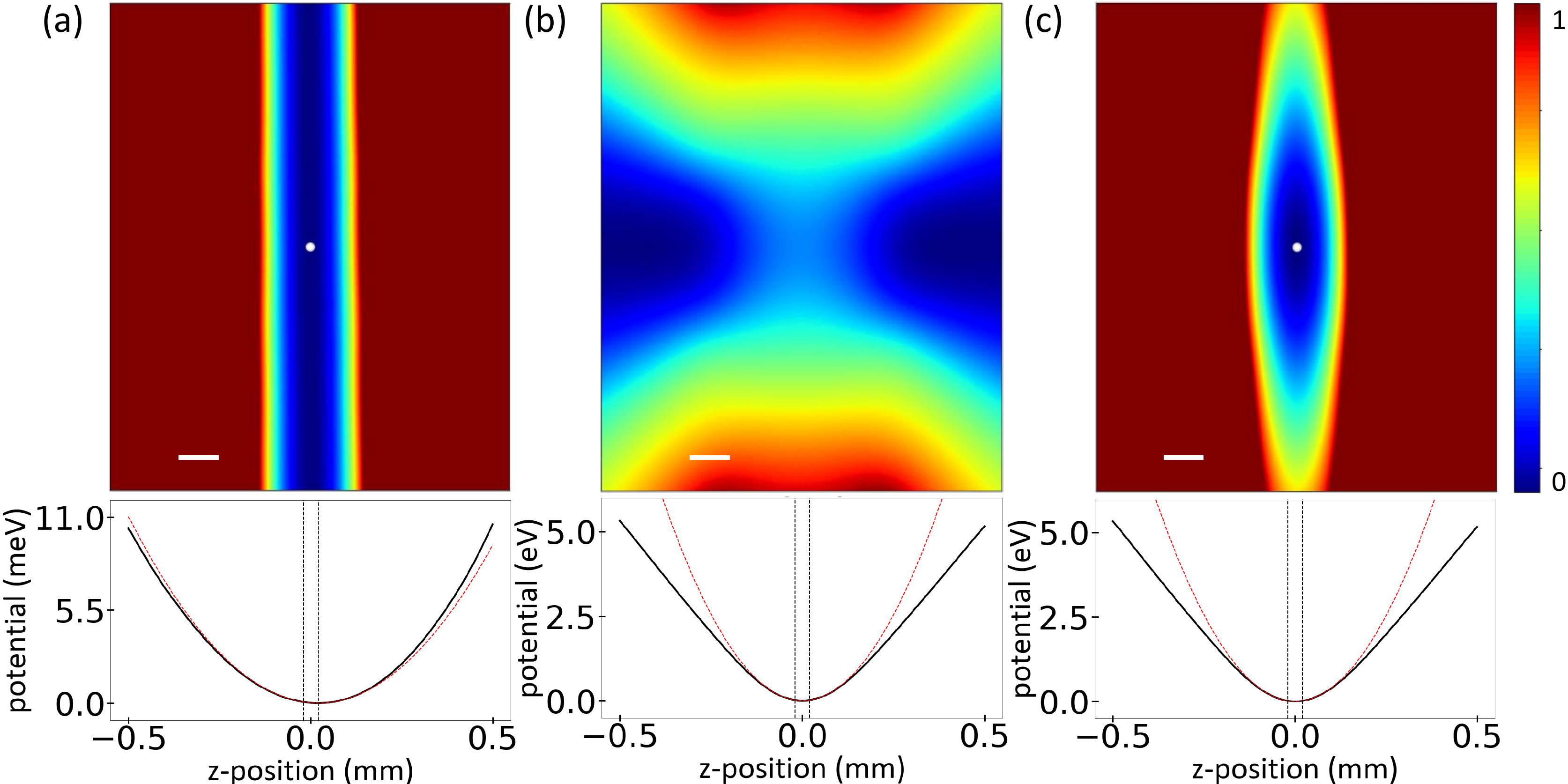}
	\caption{
	Contour plot of pseudopotential in the $yz$ plane at $x=0.2$~$\mu$m and slicecut at $y=1.6$~$\mu$m in (a), dc potential in (b), and total potential in (c).
	Scale bar denotes $100$~$\mu$m.
	Black solid line is the slicecut, red dashed line is quadratic fit, and black dashed line shows the fitting range.}
	\label{fig:axial_micromotion}
\end{figure}

We proceed to investigate the effect of the voltage change in dc electrodes to the shift of the potential minimum. 
We simultaneously add a common offset voltage $\Delta V_{\rm{dc}}$ to all the dc electrodes at $x>0, y>0$ and $-\Delta V_{\rm{dc}}$ to every dc electrode in $x<0, y<0$.
The results are plotted in Fig.~\ref{fig:micromotion}(b), showing that the voltage impression of $\Delta V_{\rm{dc}}$ moves the potential minimum along the direction of $\pm(\hat{x} + \hat{y})/\sqrt{2}$, in the central region of the trap.
The positions are moved over $\sim$10~$\mu$m as $\Delta V_{\rm{dc}}$ changes by 4~V.
From linear fittings in Fig.~\ref{fig:micromotion}(b), we obtain the slopes of 2.64(1)~$\mu$m$/\Delta V_{\rm{dc}}$ and 2.66(1)~$\mu$m$/\Delta V_{\rm{dc}}$ along the $x$ and $y$ axis, respectively. 

Armed with such understanding of the biasing rods and dc electrodes, we seek voltage sets that move the potential minimum solely along the $y$ axis.
To this end, with trials and errors, we obtain a relation between $V_{\rm{rod}}$ and $\Delta V_{\rm{dc}}$ as $V_{\rm{rod}} = -59.1 \times \Delta V_{\rm{dc}}$, which cancels the $x$-displacement by $\Delta V_{\rm{dc}}$ with $V_{\rm{rod}}$.
We remark that this value is slightly larger than the scaling of the above values as $V_{\rm{rod}} = -58.3(4) \times \Delta V_{\rm{dc}}$. 
It is because the $x$ component of an electric field by $V_{\rm{rod}}$ gently decreases as $|y|$~$\mu$m increases.
Note that Fig.~\ref{fig:micromotion}(a), from which we find the above relation of 45.3(4)~nm/$V_{\rm{rod}}$, is obtained at $y=1.4$~$\mu$m where the $y$ component of the field is close to zero (and the $x$ component of an electric field per $V_{\rm{rod}}$ is almost maximum).
However, we look for $V_{\rm{rod}}$ that removes the $x$ displacement at different $y$ coordinates, where the small $y$ component of the field appears and $x$ component decreases accordingly: as $|y|$~$\mu$m becomes larger, given the same $V_{\rm{rod}}$, the displacement along the $x$ axis (by $V_{\rm{rod}}$) decreases, which requires higher $V_{\rm{rod}}$ to cancel out the $x$ displacement by identical $\Delta V_{\rm{dc}}$. 
This small discrepancy was solved by our ``trials and errors''---we simulate many times to find the voltage ratio of $-59.1$ that suppresses the net $x$ displacement as much as possible.
Fig.~\ref{fig:micromotion}(c) presents simulation results that the $y$ position is moved over $\sim$10~$\mu$m when $\Delta V_{\rm{dc}}$ varies by $4$~V; the displacement along the $x$ axis is canceled by impressing $V_{\rm{rod}}$ simultaneously.
The linear fitting of Fig.~\ref{fig:micromotion}(c) yields the $y$-displacement per $\Delta V_{\rm{dc}}$ of 2.65(1)~$\mu$m/$\Delta V_{\rm{dc}}$ along the $y$ axis, which is very close to the value of Fig.~\ref{fig:micromotion}(b).
Combining the results along both the axes, it would be possible to locate the potential minimum at arbitrary points in the transverse plane. 
We note that all error bars in this subsection are originated from the precision of our numerical simulation.

\subsection{Blade misalignment}
\label{sec:blade_misalignment}
 
\subsubsection{Potential minimum}

Our last investigation focuses on the influence of blade misalignment to the trapping potential. 
Although all considerations above are presumed on perfect alignment of the blades, in practice, it is not possible to avoid angular or lateral misalignments of the blades. 
Such misalignment distorts the shape of trap potentials, which would cause an instability of the ions.  
Here we describe one of such misalignments, tilt between the blades. 
Fig.~\ref{fig:tilt_xy_plane}(a) presents a situation in which one rf blade ($x<0, y>0$) is tilted at an angle of $1.5^{\circ}$ with respect to the $z$ axis. 
The simulation results are shown in Figs.~\ref{fig:tilt_xy_plane}(b)--(f) when $V_{\rm{rf}}=600$~V, and other dc electrodes and biasing rods are ground.
In case of $z<0$, the rf blade is located more distantly from the origin with respect to other blades. 
Therefore, the effect of rf voltages in this blade is reduced, moving the rf null position toward the region of $x<0, y>0$ (Figs.~\ref{fig:tilt_xy_plane}(b), (c)). 
By contrast, for the regime of $z>0$, this tendency appears in an opposite way so that the rf null is located at $x>0, y<0$ (Figs.~\ref{fig:tilt_xy_plane}(e), (f)).
Fig.~\ref{fig:tilt_xy_plane}(g) shows the calculation results of rf nulls along the trap axis. 
We also present both the cross section in the $yz$ plane and slicecut along the $z$ direction as the $x$ coordinate changes (Figs.~\ref{fig:tilt_yz_plane}(c)--(g)).
The potential is very asymmetric along the $z$ axis, and the trap depth is overall larger in the region of $z>0$ than that of $z<0$.

\subsubsection{Axial micromotion}

We point out that this kind of blade rotation causes an axial micromotion~\cite{Herschbach2012, Low2019}, which would be very difficult to be compensated. 
In our consideration, whereas the axis of $\phi(\textbf{r})_{\rm{dc}}$ is in parallel with the $z$ axis, the axis of the rf tube is not: In consequence, the axis of $\phi(\textbf{r})_{\rm{tot}}$, along which the ions aligned, is not parallel with the axis of $\phi(\textbf{r})_{\rm{rf}}$ --- this induces an inevitable axial micromotion.
Therefore, it is crucial to align the blades as parallel as possible over the course of actual trap development.

We quantify the axial micomotion in this tilted configuration. 
With the same rf voltage and dc voltage set given in Sec.~\ref{sec:confinement_in_three_dimensions}, our calculation of $\phi(\textbf{r})_{\rm{tot}}$ finds the potential minimum at $\mathbf{\bar{r}} = (0.2, 1.6, 0.5)$~$\mu$m (Fig.~\ref{fig:axial_micromotion}), at which the ion is located. 
We note that, under perfect alignment, the potential minimum is found at (0.2, 1.6, 0.0)~$\mu$m in Sec.~\ref{sec:confinement_in_three_dimensions}: this $z$-position shift is induced by the tilt misalignment, and the rf field at $\mathbf{\bar{r}}$ along the $z$ direction causes the axial micromotion. 
Our numerical simulation provides an amplitude of the rf field $E_{\rm{rf, z}}(\mathbf{\bar{r}})=59.8$~V/m along the $z$ axis with an equivalent energy of $\phi(\mathbf{\bar{r}})_{\rm{rf}}=25.2$~neV.
This energy corresponds to an ``excess'' temperature of $T_{\rm{axial}}=585$~$\mu$K.

\section{Discussion}

We remark four points regarding the present study. 
First, we describe how we choose the location of the rods. 
The coordinates of the two rods' centers are $(x, y)=(-2.37, \pm1.81)$~mm, determined by two practical reasons. 
One point is that, in a given position, it should be possible to move the rf-null over tens of microns in hundreds of volts, in which we can apply using a commercially available power supply. 
Another reason consists of the optical access: 
Along this direction, our numerical aperture (NA) is 0.40 limited by a recessed viewport of our vacuum chamber.
Therefore we aim to attain a NA governed by the rods larger than 0.40, which does not influence imaging of the ions.

Next, during the investigation of micromotion compensation in Sec.~\ref{sec:micromotion}, one might think that the $y$-position shift would be possible by installing two biasing rods at the area $y>0$. 
We separately calculate the shift of the potential minimum in such configuration, however, the displacement is very small that the position only moves by $\sim0.2$~$\mu$m in 1000~V.
It is because, the electric fields from biasing rods at $y>0$ are more shielded by the blades than the fields from the rods at $x<0$, due to the asymmetric angles between the blades.

Third, from the perspective of quantum networks and distributed quantum computing, optical interface with the segmented-blade trap was mostly done with high-NA lenses so far~\cite{Hucul2015, Craddock2019, Siverns2019}. 
Alternatively, we point out that it would also be interesting to combine the trap with optical cavities~\cite{Lee2019, Walker2020, Meraner2020}.
It would be informative if future theoretical studies could include a strategy for coupling fiber cavities~\cite{Steiner2013, Brandstaetter2013, Takahashi2017, Lee2019a}, the influence of dielectric surface charges to the trap potential~\cite{Ong2020}, and the impact of high voltage on piezoelectric transducer (for scanning the cavity length) to the potential energy.

Lastly, while we herein concentrate on the calculation of the trap potential, the vacuum pressure, particularly the local pressure near the ions, is also an important factor for stable ion trapping. 
It is notable that the software MolFlow+~\cite{kersevan1991} could be used for estimating the local pressure at the location of the ions. 
By doing so, we expect that the trap mount and other vacuum components could be designed and installed in an optimized geometry: the system should be designed such that the conductance from the ion position to the vacuum pumps is maximized.

\section{Conclusion}

In conclusion, we have presented a numerical investigation of a linear ion trap, including four segmented blades and two biasing rods. 
We find that it is possible to achieve an excellent optical access for the radial directions. 
Also, the trap depth and frequency are calculated in a given voltage set, which offers a stable trapping configuration of $^{171}$Yb$^{+}$ ions.
We then describe an approach to compensate the micromotion, followed by a calculation regarding the blade misalignment.
Our study provides thorough understanding of this ion-trapping system, forming the basis of stable ion-based quantum computer and simulator.

\section{acknowledgments}

We thank Kihwan Kim for helpful discussions. 
This work has been supported by National Research Foundation (Grant No.~2019R1A5A1027055), Institute for Information and Communication Technology Planning and Evaluation (IITP, Grant No.~2022-0-01040), Samsung Science and Technology Foundation (SSTF-BA2101-07, SRFC-TC2103-01), Samsung Electronics Co., Ltd.~(IO201211-08121-01), and BK21 FOUR program.

\bibliographystyle{apsrev4-2}
\bibliography{bibliography}

\clearpage
\twocolumngrid

\appendix

\section{rf-null position shift}
\begin{figure} [h]
	\includegraphics[width=3.3in]{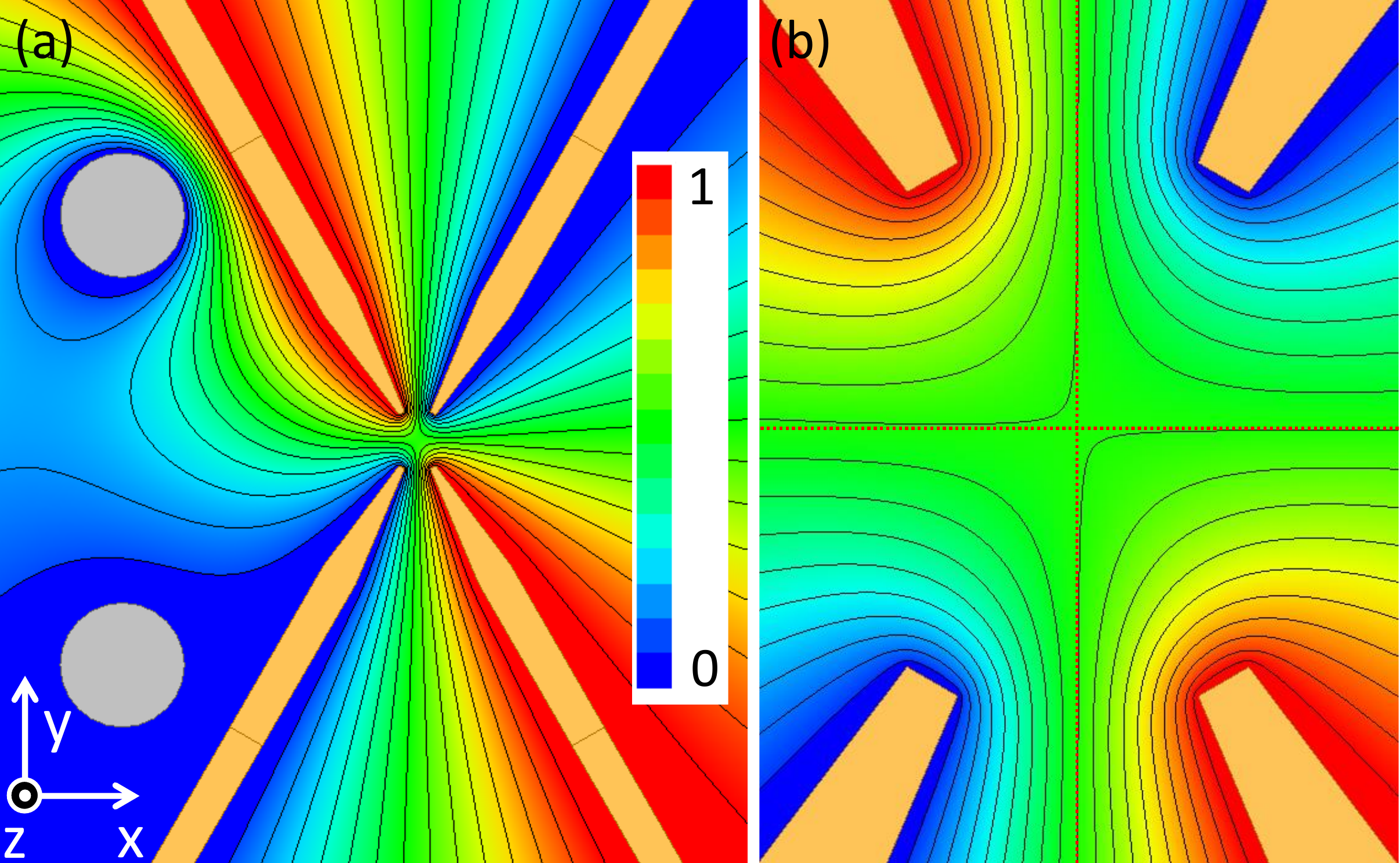}
	\caption{
	(a) Contour plot of rf quadrupole potential with the parameters used in Figs.~\ref{fig:pseudopotential}(a)--(f).
	(b) Central area. 
	Red dashed lines are $x$ and $y$ axes.
	It is notable that rf null is shifted to $+y$ direction slightly. 
	}	
	\label{quadrupole}
\end{figure}

\section{Blade mismatch along the radial direction}

\begin{figure*} [h]
	\includegraphics[width=6.3in]{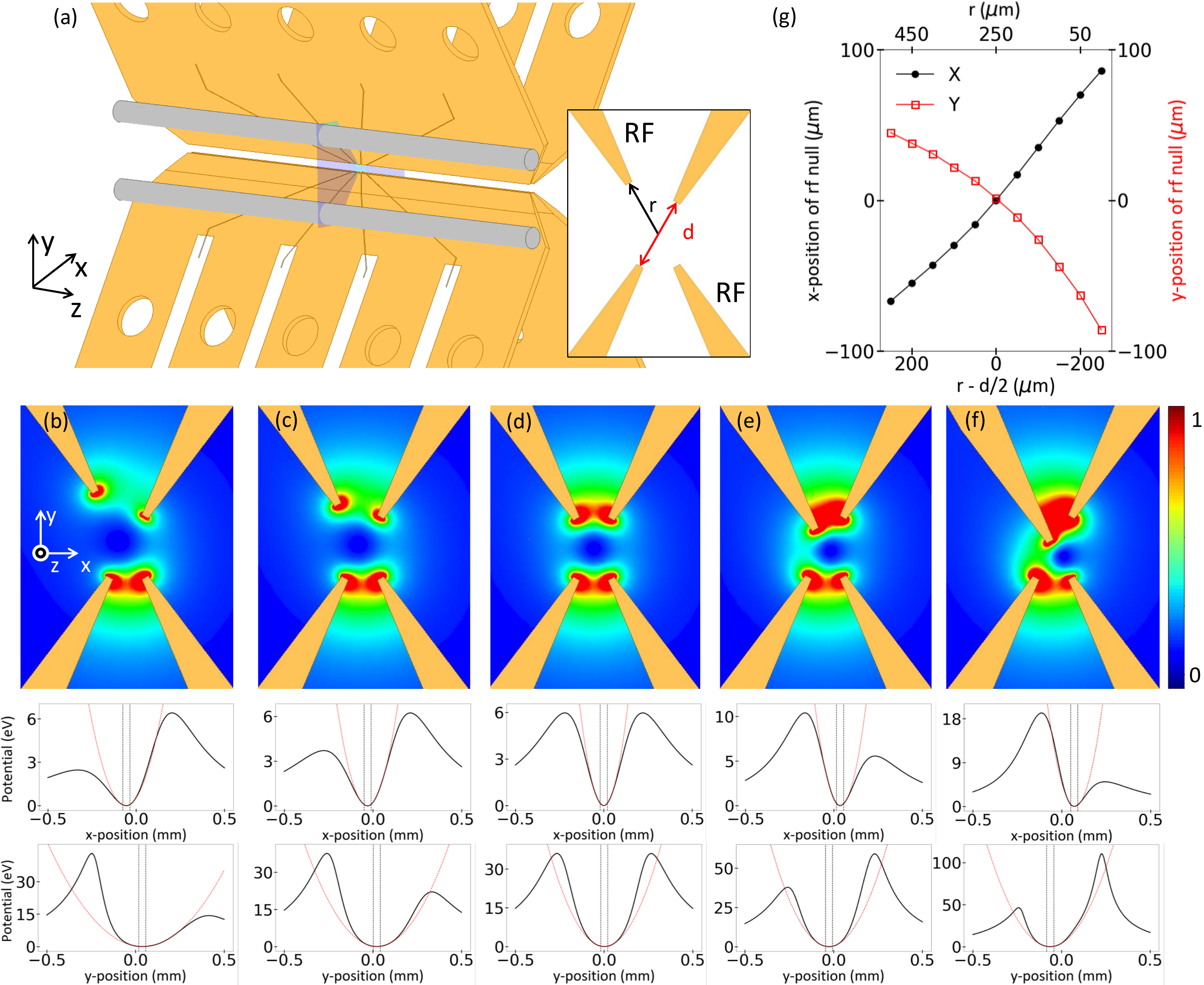}
	\caption{
	        (a) Rf blade at $x<0, y>0$ is translated transversally by a distance $r-d/2$. 
	        Blade-blade distance is $d=500$~$\mu$m.
		Semi-transparent plane denotes the position of cross sections shown in (b)--(f).
		Contour plot of pseudopotential at $r-d/2=200$~$\mu$m in (b), $r-d/2=100$~$\mu$m in (c), $r-d/2=0$~$\mu$m in (d), $r-d/2=-100$~$\mu$m in (e), and $r-d/2=-200$~$\mu$m in (f). 
		(g) Coordinates of rf null are plotted as functions of $r-d/2$ and $r$.
	}	
	\label{fig:mismatch_radial}
\end{figure*}

We investigate the influence of a mismatch in which one of the rf blades is transversally displaced from the original position. 
Fig.~\ref{fig:mismatch_radial}(a) shows that the rf blade at $x<0$ and $y>0$ is translated by $r-d/2$. 
We numerically calculate $\phi(\textbf{r})_{\rm{rf}}$ as $r-d/2$ changes from $200$~$\mu$m to $-200$~$\mu$m, show the results in Figs.~\ref{fig:mismatch_radial}(b)--(f), and find rf-null points in Fig.~\ref{fig:mismatch_radial}(g).
When $r$ is larger than the original position $d/2$, the rf null moves toward the position of $x<0$ and $y>0$, and the tendency becomes opposite at $r<d/2$.

We remark the similarity/difference between this transversal misalignment and the tilt misalignment in Sec.~\ref{sec:blade_misalignment}.
In Fig.~\ref{fig:tilt_xy_plane}, the cross section of $\phi(\textbf{r})_{\rm{rf}}$ and the rf-null positions are plotted as the $z$ coordinate varies.
The behavior in a plane of Fig.~\ref{fig:tilt_xy_plane}(a) is very close to one result of Fig.~\ref{fig:mismatch_radial}.
Given $z$ position where the distance between the rf blades is larger than $d$, the potential shape and shift of the rf null is similar with the results at $r-d/2>0$ of Fig.~\ref{fig:mismatch_radial}; when the distance between the rf blades is smaller than $d$ (Fig.~\ref{fig:tilt_xy_plane}), the similarity is found with the result at $r-d/2<0$ (Fig.~\ref{fig:mismatch_radial}).
Regarding the difference, in this transversal mismatch, the rf tube and trap axis (defined by $\phi(\textbf{r})_{\rm{tot}}$) run parallel, which does not cause an axial micromotion.

\end{document}